**Title: The relationships between PM$_{2.5}$ and AOD in China: About and behind spatiotemporal variations**

**Authors:** Qianqian Yang [a], Qiangqiang Yuan[*, a], Linwei Yue[*, b], Tongwen Li [b], Huanfeng Shen [b], Liangpei Zhang [c]

**Affiliations:**

[a] School of Geodesy and Geomatics, Wuhan University, Wuhan, Hubei, 430079, China.

[b] Faculty of Information Engineering, China University of Geosciences, Wuhan, Hubei, 430074, China.

[c] School of Resource and Environmental Sciences, Wuhan University, Wuhan, Hubei, 430079, China

[d] State Key Laboratory of Information Engineering, Survey Mapping and Remote Sensing, Wuhan University, Wuhan, Hubei, 430079, China

[*] **Corresponding author:**

Qiangqiang Yuan (yqiang86@gmail.com)
Linwei Yue (yuelw@cug.edu.cn)

**ABSTRACT**

Satellite aerosol products have been widely used to retrieve ground $PM_{2.5}$ concentration because of its wide coverage and continuously spatial distribution. While more and more studies focus on the retrieval algorithm, we find that the relationship between $PM_{2.5}$ concentration and satellite AOD has not been fully discussed in China. Is satellite AOD always a good indicator for $PM_{2.5}$ in different regions and can AOD still be employed to retrieve $PM_{2.5}$ with pollution conditions changing in these years remain unclear. In this study, the relationships between $PM_{2.5}$ and AOD were investigated in 368 cities in China for a continuous period from February 2013 to December 2017, at different time and regional scales. Pearson correlation coefficients and $PM_{2.5}$/AOD ratio were used as the indicator. Firstly, we concluded the relationship of $PM_{2.5}$ and AOD in terms of spatiotemporal variations. Then the impact of meteorological factors, aerosol size and topography were discussed. Finally, a GWR retrieval experiment was conducted to find out how was the retrieval accuracy changing with the varying of $PM_{2.5}$-AOD relationship. We found that spatially the correlation is higher in Beijing-Tianjin-Hebei and Chengyu region and weaker in coastal areas such as Yangtze River Delta and Pearl River Delta. The $PM_{2.5}$/AOD ratio has obvious North-South difference with a high ratio in north China and a lower ratio in south China. Temporally, $PM_{2.5}$/AOD ratio is higher in winter and lower in summer, the correlation coefficient tends to be higher in May and September. As for interannual variations from 2013 to 2017, we detected a declining tendency on $PM_{2.5}$/AOD ratio. The accuracy of GWR retrievals were decreasing too, which may imply that AOD may not be a good indicator for $PM_{2.5}$ in the future. Our study firstly investigates the $PM_{2.5}$-AOD relationship in such a large extent at city scale, and firstly investigate the temporal variations in terms of interannual variations. The results will be very helpful for $PM_{2.5}$ concentration satellite retrieval and can help us further understand the $PM_{2.5}$ pollution in China.

# 1. Introduction

Particulate matter (PM) is the fraction of air pollution which is most reliably associated with human disease(Anderson et al., 2011). Exposure to particulates especially fine particulate matter with an aerodynamic diameter of less than 2.5 $\mu$m ( $PM_{2.5}$) can have serious adverse impact on human health(Burnett et al., 2014; Harrison et al., 2012; J. Lelieveld, 2015). China has experienced both rapid urbanization and serious $PM_{2.5}$ pollution in the past decades. It was reported that more than 1.35 million people are killed by air pollution each year, and the contribution of $PM_{2.5}$ was higher than 95% (J. Lelieveld, 2015). The huge loss brought by $PM_{2.5}$ pollution has attracted the government's attention on $PM_{2.5}$ concentration monitoring issues.

In recent years, retrieving surface $PM_{2.5}$ concentration through satellite aerosol products has become a popular $PM_{2.5}$ monitoring method due to its wide coverage and continuous spatial distribution, aerosol optical depth (AOD) is the most commonly used satellite product. At the very beginning, Wang et al. (Wang, 2003) compared the surface $PM_{2.5}$ concentration with satellite AOD first and found a high correlation with R=0.7 in 2003 in a county of America. Then they built a linear equation for 24-h mean $PM_{2.5}$ and MODIS AOD and proved that $PM_{2.5}$ derived from the MODIS AOD can be quantitatively used to estimate the air quality categories. Since then, many scholars started to retrieve surface $PM_{2.5}$ concentration from AOD in different countries and regions with different retrieving techniques (de Hoogh et al., 2018; Fang et al., 2016; He and Huang, 2018; Jung et al., 2017; Li et al., 2017a; Li et al., 2017b; Tian and Chen, 2010; van Donkelaar et al., 2015; Yin et al., 2016). These techniques mainly aim at building a mapping relation between $PM_{2.5}$ concentration and AOD, the mapping function can be empirical, statistical, or semi-empirical. Empirical methods concentrate on figuring out the physical connection between column AOD and surface PM$_{2.5}$ concentration and build the mapping function through conducting humidity correction, vertical correction etc. Statistical methods don't care about the physical connection between PM$_{2.5}$ and AOD, and just using statistical models such as Multiple Linear Regression (MLR), Linear Mixed Effect Model (LME) and Artificial Neural Networks (ANN) to build the mapping function. And the last semi-empirical model is a combination of the above two methods.

In fact, PM$_{2.5}$ retrieval methods have developed a lot in recent decades, from univariate to multivariate and from linear to nonlinear. An important premise and theoretical foundation for retrieving surface PM$_{2.5}$ concentration with satellite AOD is the strong correlation and connection between PM$_{2.5}$ and AOD. Ground surface PM$_{2.5}$ concentration and AOD both represent part of the suspended matter in the atmosphere and can both reflect the turbidity of atmosphere to a degree, so there is possibility that they are highly correlated and physically connected. But there are also many factors that may weaken the correlation between measured PM$_{2.5}$ concentration and satellite AOD. For example, PM$_{2.5}$ mainly represents the turbidity of atmosphere near the ground. But AOD measures the whole atmospheric column which extends from ground surface to an altitude of several hundred kilometers. Besides, PM$_{2.5}$ mainly represents the dry mass concentration of fine particulate which is hardly affected by water and coarse particles, but the value of AOD include the influence of water vapor and coarse particles. Apart from this, in essence, the value of PM$_{2.5}$ stands for mass concentration, but AOD value represents the extinction ability, the connection between mass concentration and extinction ability can be either strong or weak with the composition of PM$_{2.5}$ and aerosol varying. To sum up, the relationship between PM$_{2.5}$ concentration and AOD is complicated. In different conditions, these factors may weaken the relationship to a different degree, a high correlation in a region during a period cannot represent a constant high correlation in all regions and all periods because the environments and conditions may change. That's to say, when retrieving $PM_{2.5}$ concentration through AOD in a large area for a long time series, it's necessary to explore the relationship between $PM_{2.5}$ and AOD for the same large spatial and temporal extent, and find out the spatial and temporal variations of the relationship at a fine scale. Only by doing this, can we figure out whether the foundation is always solid when retrieving $PM_{2.5}$ with AOD in whole area for a long time.

However, at present, we can find a lot of studies researching the long time series retrieval for $PM_{2.5}$ in whole China(Boys et al., 2014; Ma et al., 2016b), but we can hardly find a paper exploring the relationship between $PM_{2.5}$ and AOD for whole China for a long time series. Most of current studies about $PM_{2.5}$-AOD relationship in China are confined to a certain city or a small area. For example, Shao et al.(Shao et al., 2017) studied the correlation between $PM_{2.5}$ and AOD in Nanjing and find a strong correlation with $R^2=0.56$; Similar work was conducted in Beijing-Tianjin-Hebei and the correlation coefficients r varies from 0.52 to 0.79 (Ma et al., 2016a); Wang et al. studied the relationship in Chongqing, Sichuan and the correlation between $PM_{2.5}$ concentration and AOD is 0.702(Wang et al., 2018). Studies for single region can be useful in the specific area but usually cannot apply to other regions, therefore help little in the large-scale retrieval problem. Furthermore, there were a few studies investigating the $PM_{2.5}$-AOD relationship in a larger spatial extent and discuss the spatial variations. Guo et al.(Guo et al., 2009) studied the correlation between PM concentration and AOD in 11 PM observation sites from China Atmosphere Watch Network (CAWNET) in eastern China in 2009 and emphasize the importance of spatial variations of PM-AOD relationship with foresights. In their later work in 2017(Guo et al., 2017a), they studied the spatial variations of $PM_{2.5}$-AOD relationship over the whole China based on 50km×50km grids and find a strong correlation in North China Plain(0.50<R<0.71) and relatively weaker correlation in PRD and YRD(0.35<R<0.55). But the resolution is quite coarse and some important spatial variation details may have been ignored. Xin et al. (Jinyuan Xin1 and Yang Sun1, 2016) also compared the correlation between $PM_{2.5}$ concentration and AOD in different locations, but the AOD data used in experiments are from Campaign on atmospheric Aerosol REsearch network of China(CARE-China) which is measured by ground sites rather than satellite sensors. Besides, the sites distribution of this network is sparse and only 23 sites are analyzed. Given all that, current $PM_{2.5}$-AOD relationship analysis work are either confined to a small area or not confined to a small area but lacks investigation about detailed spatial variations. Hence, to understand the $PM_{2.5}$-AOD relationship over China more comprehensively and thoroughly, study for large area with comparisons for different cities and regions is needed.

In addition to spatial variations, temporal variations of $PM_{2.5}$-AOD relationship also worth more attention. According to some recent studies, the $PM_{2.5}$ pollution in China has been eased in recent years with government's policy taking effect(Lin et al., 2018). With the air quality getting better, the physical and chemical characteristics, the horizonal and vertical distribution feature of $PM_{2.5}$ and AOD may also have changed a lot, which may lead to the change of $PM_{2.5}$-AOD relationship. Furthermore, the change of $PM_{2.5}$-AOD relationship may bring about the change of retrieval performance as well. Previous studies researching the temporal variations of $PM_{2.5}$-AOD relationship mainly focus on the seasonal variation(Li et al., 2015; Ma et al., 2016a), and interannual variations has been ignored. Hence, how is the relationship changing in recent years worth further researching(Li et al., 2015).

Based on the above two points, our study aimed at comprehensively investigating the relationships between $PM_{2.5}$ concentration and satellite AOD in China, with an emphasis on spatial distribution pattern and temporal variations especially interannual variations. We explore the $PM_{2.5}$-AOD relationship in whole China at city level and investigate the temporal variations in terms of interannual variations based on a comparatively long time series analysis. In this study, the relationship analysis was conduct in 368 cities and 9 metropolitan regions in China based on a 59-month record of observations from February 2013 to December 2017. The relationship between $PM_{2.5}$ concentration and AOD was measured by the Pearson correlation coefficients between them and the $PM_{2.5}$/AOD ratio. We compared the spatial difference of the $PM_{2.5}$-AOD relationship at city and region scale. The monthly and interannual variations were also explored. Then we discussed the impact of some influencing factors including planetary boundary layer height (PBLH), relative humidity (RH), fine mode fraction (FMF), Angstrom exponent (AE) and topography for $PM_{2.5}$-AOD relationship. Finally, a simple geographical weighted regression (GWR) retrieval was conducted to validate $PM_{2.5}$-AOD relationship's impact on retrieval accuracy and find out the interannual variations of retrieval performance. The conclusions of this study could provide some useful instructions for $PM_{2.5}$ retrieval, for example, the spatial and temporal varying pattern of $PM_{2.5}$-AOD relationship could provide a reference for the development of spatially and temporally self-adaptative retrieval algorithm. Besides, the discussion of impacting factors for $PM_{2.5}$-AOD relationship can improve our understanding of the formation mechanisms of air pollution.

The rest of this paper is organized as follows. Section 2 is the data and method part, where we introduce the study area and period, the data sources, the data preprocessing work, and the methodology, and provide a flow chart of the study design. The experimental results and a discussion are provided in Section 3. Finally, we make a summary of our work in Section 4.

## 2. Data and method

*2.1. Study area and period*

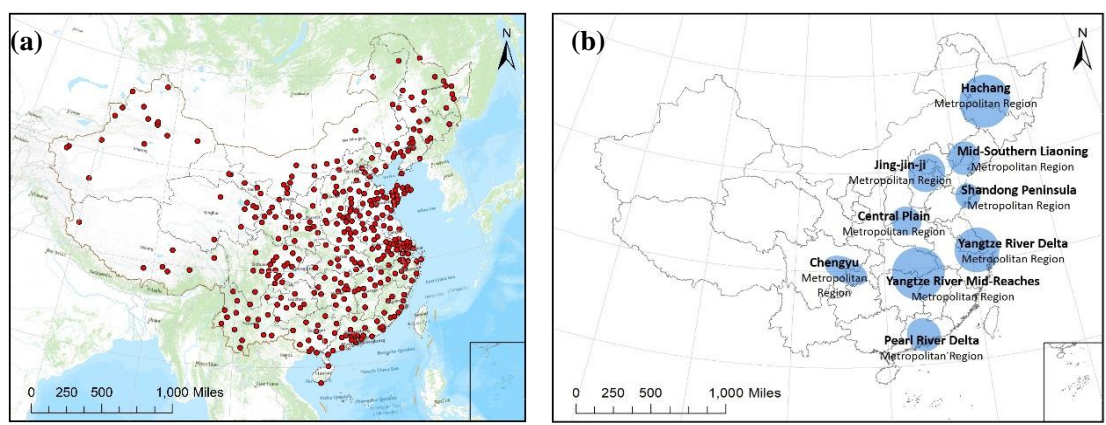

**Fig. 1.** Study area:(a) the 368 cities and (b) 9 metropolitan regions in China.

China has experienced serious air pollution in recent decades. Our study covered 368 cities and 9 metropolitan regions in China as shown in Figure.1. Although most of the environmental monitoring sites are located in economically developed areas such as Jing-jin-ji Metropolitan Region (BTH), Yangtze River Delta (YRD) and Pearl River Delta (PRD), the distribution of 368 cities still spread over all of China and there is at least one city studied for each province, making the analysis of spatial variations feasible. The 9 metropolitan regions include the Jing-jin-ji Metropolitan Region(BTH), Yangtze River Delta Metropolitan Region(YRD), Pearl River Delta Metropolitan Region (PRD), Central Plain Metropolitan Region(CP), Chengyu Metropolitan Region(CY), Yangtze River Mid-Reaches Metropolitan Region(YRM), Shandong Peninsula Metropolitan Region(SP), Mid-Southern Liaoning Metropolitan

Region(MSL) and Hachang Metropolitan Region(HC), all of which are national metropolitan regions approved by the National Assembly of China. These 9 national metropolitan regions are densely populated and economically developed areas in China and have an important position in the development of national strategy. Therefore, we choose them to be the object of regional scale study. The detailed information for the 9 metropolitan regions is listed in Table.1.

The time range of our study was from February 3, 2013, to December 31, 2017, i.e., nearly 59 months and 5 years in total.

Table 1. The detailed information for 9 metropolitan regions in our study.

| Num | Metropolitan Region | Abbreviation | Total Area ($\times 10^4 km^2$) | Cities included (number of cities) |
|---|---|---|---|---|
| 1 | Jing-jin-ji Metropolitan Region | BTH | 22.6 | Beijing, Tianjin, Tangshan, Baoding, Shijiazhuang, Qinhuangdao, Langfang, Zhangjiakou, Chengde, Cangzhou, Xingtai, Handan, Hengshui (13) |
| 2 | Yangtze River Delta Metropolitan Region | YRD | 21.2 | Shanghai, Nanjing, Wuxi, Changzhou, Suzhou, Nantong, Yancheng, Yangzhou, Zhenjiang, Taizhou, Hangzhou, Ningbo, Jiaxing, Huzhpu, Shaoxing, Jinhua, Zhoushan, Taizhou, Hefei, Wuhu, Ma'anshan, Tongling, Anqing, Chuzhou, Chizhou, Xuancheng (26) |
| 3 | Pearl River Delta Metropolitan Region | PRD | 42.2 | Guangzhou, Shenzhen, Zhuhai, Foshan, Dongguan, Huizhou, Zhongshan, Jiangmen, Zhaoqing, Sahnwei, Qingyuan, Yunfu, Heyuan, Shaoguan (14) |
| 4 | Central Plain Metropolitan Region | CP | 28.7 | Zhengzhou, Luoyang, Kaifeng, Nanyang, An'yang, Shangqiu, Xinxiang, Pingdingshan, Xuchang, Jiaozuo, Zhoukou, Xinyang, Zhumadian, Hebi, Puyang, Luohe, Sanmenxia, Jiyuan, Changzhi, Jincheng, Yuncheng, Liaocheng, Heze, Suzhou, Huaibei, Fuyang, Bengbu, Haozhou, Xingtai, Handan (30) |
| 5 | Chengyu Metropolitan Region | CY | 18.5 | Chngqing, Chengdu, Zigong, Luzhou, Deyang, Mianyang, Suining, Neijiang, Leshan, Nanchong, Meishan, Yibin, Guang'an, Dazhou, Ya'an, Ziyang (16) |
| 6 | Yangtze River Mid-Reaches Metropolitan Region | YRM | 31.7 | Wuhan, Huangshi, Ezhou, Huanggang, Xiaogan, Xianning, Xiantao, Qianjiang, Tianmen, Xiangyang, Yichang, Jinmen, Jinzhou, Changsha, Zhuzhou, Xiangtan, Yueyang, Yiyang, Changde, Hengyang, Loudi, Nanchang, Jiujiang, Jingdezhen, Yingtan, Xinyu, Yichun, Pingxiang, Sahngrao, Fuzhou, Ji'an (31) |
| 7 | Shandong Peninsula Metropolitan Region | SP | 7.3 | Jinan, Qingdao, Yantai, Zibo, Linyi, Weifang, Dongying, Weihai, Rizhao, Zaozhuang, Heze, Jining, Tai'an, Laiwu, Binzhou, Dezhou, Liaocheng (17) |

| 8 | Mid-Southern Liaoning Metropolitan Region | MSL | 9.7 | Shenyang, Dalian, Anshan, Fushun, Benxi, Dandong, Liaoyang, Yingkou, Panjin (9) |
| 9 | Hachang Metropolitan Region | HC | 26.4 | Harbin, Daqing, Qiqihar, Suihua, Mudanjiang, Changchun, Jilin, Siping, Liaoyuan, Songyuan, Yanbian (11) |

*2.2. Data collection*

*2.2.1 $PM_{2.5}$ concentration data*

Since 2012, China has started building environmental monitoring sites all over the country, and the concentration of 6 main atmospheric pollutants including $PM_{2.5}$, $PM_{10}$, $SO_2$, $NO_2$, $O_3$, CO have been published online since January 2013. We downloaded the hourly $PM_{2.5}$ concentration data from the Data Center of the Ministry of Environmental Protection of the People's Republic of China (http://datacenter.mep.gov.cn/index), from 2013 to 2017. By December 2017, the number of $PM_{2.5}$ monitoring station had approached 1500 and spread over 368 cities. The $PM_{2.5}$ concentration is measured by Tapered Element Oscillating Microbalance (TEOM) or beta attenuation monitor. Daily average $PM_{2.5}$ concentration is averaged from hourly values.

*2.2.2 Satellite AOD data*

The satellite AOD product is provided by the Moderate Resolution Imaging Spectroradiometer (MODIS) on Terra. The MODIS instrument is a multi-spectral radiometer, designed to retrieval aerosol microphysical and optical properties over land and ocean(Levy et al., 2013; Tanré et al., 1997). The 2330 km swath width of the MODIS instrument produces a global coverage in 1 or 2 days and captures most of aerosol variability due to this high sampling frequency. In this study, the MODIS level 2 daily AOD data from Terra(MOD04_L2, Collection 6) are used, which is reported at 10km×10km, and are within uncertainty levels of $\pm 0.05 \pm 0.20 \times$ AOD over land(Chu, 2002; Levy et al., 2007). We used the 10km AOD product instead of 3km higher resolution product since the data missing problem is less serious in the 10km resolution product. The MODIS AOD data product was downloaded from NASA LAADS (https://ladsweb.modaps.eosdis.nasa.gov/).

*2.2.3 Meteorological data*

Meteorological data used in our study is downloaded from NCEP/NCAR Reanalysis1: Surface dataset (https://www.esrl.noaa.gov/psd/data/gridded/data.ncep.reanalysis.surface.html) and five meteorological parameters are downloaded from the above website: relative humidity(RH), air temperature(TMP), U-wind(UW), V-wind(VW) and pressure(PS). All the meteorological data are daily surface data. But there is no information about planetary boundary layer height (PBLH) in this data collection, so we also downloaded the PBLH data from MERRA-2. The Modern-Era Retrospective analysis for Research and Applications version 2 (MERRA-2) is a NASA atmospheric reanalysis for the satellite era using the Goddard Earth Observing System Model, Version 5 (GEOS-5) with its Atmospheric Data Assimilation System (ADAS), version 5.12.4. The PBLH data was downloaded from the collection M2T1NXFLX (https://disc.gsfc.nasa.gov/datasets/M2T1NXFLX_V5.12.4).

*2.2.4 AERONET data*

In order to acquire some information about aerosol properties, AERONET ground-based measurements are employed in our study. AERONET(Holben, 1998) is a ground-based sun-photometer network with over 1100 stations globally. The AERONET AOD and aerosol properties are derived from direct beam solar measurements(Holben et al., 2001) at two UV and five visible channels. The data used are the Version 2 Level 2 quality assured and cloud screened product, when level 2 product is not available, level 1.5 product is used for substitution.

Original data is averaged to daily and monthly data when being analyzed. Two aerosol parameters are used in our experiments: Fine mode fraction (FMF) and Angstrom Exponent (AE). There are 60 sites located in China, but most of them had stopped working for a long time and no data during our research time can be acquired. We finally choose five sites considering the sites distribution and data access. The product can be downloaded from https://aeronet.gsfc.nasa.gov/cgi-bin/type_piece_of_map_opera_v2_new.

*2.3. Methodology*

The process of our study design can be divided into five parts. The first step was the data preprocessing, including data match, outlier rejection, data average for multiscale analysis etc. Then we conducted a simple analysis of the spatiotemporal distribution pattern of PM$_{2.5}$ concentration. ArcGIS and the empirical Bayesian kriging (EBK) interpolation method (Krivoruchko et al., 2012) were used to acquire the spatial distribution map. Thirdly, we analyzed the relationship between PM$_{2.5}$ concentration and AOD at multiple scales, i.e., city, metropolitan region, month, and year scales. The relationship between daily PM$_{2.5}$ concentration and AOD was measured by the Pearson Correlation Coefficient (r) and PM$_{2.5}$/AOD ratio ($\eta$) expressed as:

$$r = \frac{\sum (X_i - \bar{X})(Y_i - \bar{Y})}{\sqrt{\sum (X_i - \bar{X})^2 \sum (Y_i - \bar{Y})^2}} \quad (1)$$

$$\eta = \frac{PM_{2.5}}{AOD} \quad (2)$$

The ratio between PM$_{2.5}$ concentration and AOD was first promoted by van Donkelaar et al. in 2010 as a conversion factor, the parameter $\eta$ (unit: $\mu g/m^3$) indicates the mass concentration of PM$_{2.5}$ per unit aerosol optical thickness (van Donkelaar et al., 2010). Previous work has proved that the PM$_{2.5}$/AOD ratio is a good parameter to measure the relationship between PM$_{2.5}$ concentration and AOD(Zheng et al., 2017). Hence, we introduce this parameter in our work in addition to correlation coefficient for a more comprehensive analysis about the PM$_{2.5}$ and AOD relationship. While the correlation coefficient r representing the synchronicity between PM$_{2.5}$ variations and AOD variations, the ratio $\eta$ reflects PM$_{2.5}$'s contribution to aerosol extinction ability.

We conducted the correlation analyses and calculated the PM$_{2.5}$/AOD ratio in both the 368 cities and 9 metropolitan regions using all the data of 5 years from 2013 to 2017, to explore the spatial variations of the relationship at different spatial scales. Then the relationship analysis for the 12 months and each year from 2013 to 2017 were implemented for detecting the monthly and interannual variations. After all this preprocessing and calculation work, we analyzed and concluded the temporal and spatial variations of the relationships between PM$_{2.5}$ concentration and AOD. To explain and understanding the spatiotemporal variations better, we also discussed the impact of some influencing factors on PM$_{2.5}$-AOD relationship. Finally, a simple PM$_{2.5}$ retrieval experiments were conducted to investigate the variations of retrieval accuracy, and to find out PM$_{2.5}$-AOD relationship's impact on retrieval accuracy. Fig. 2 shows the flow chart of the study design.

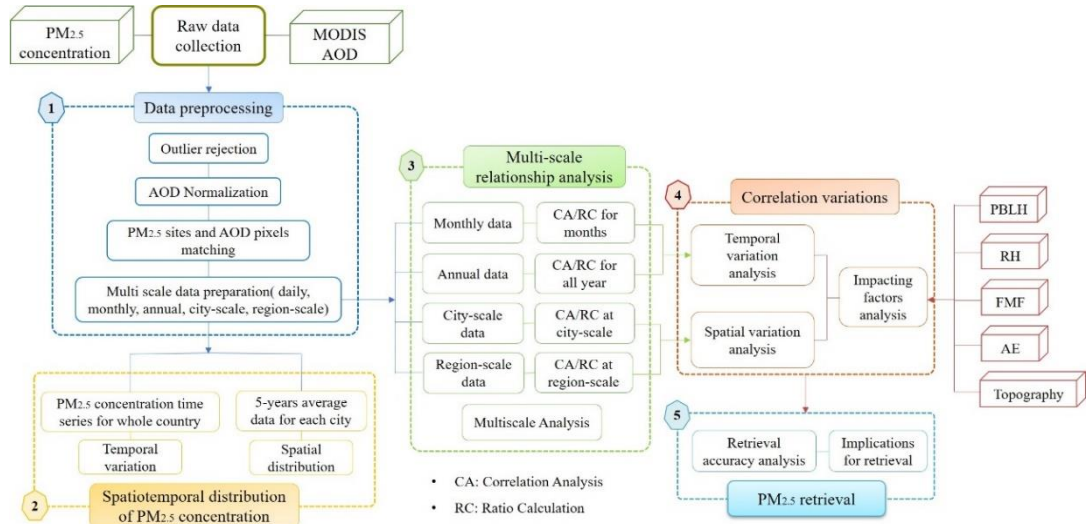

**Fig. 2.** Flow chart of the study design.

## 3. Results and discussion

*3.1. Spatiotemporal variation of PM$_{2.5}$ concentration*

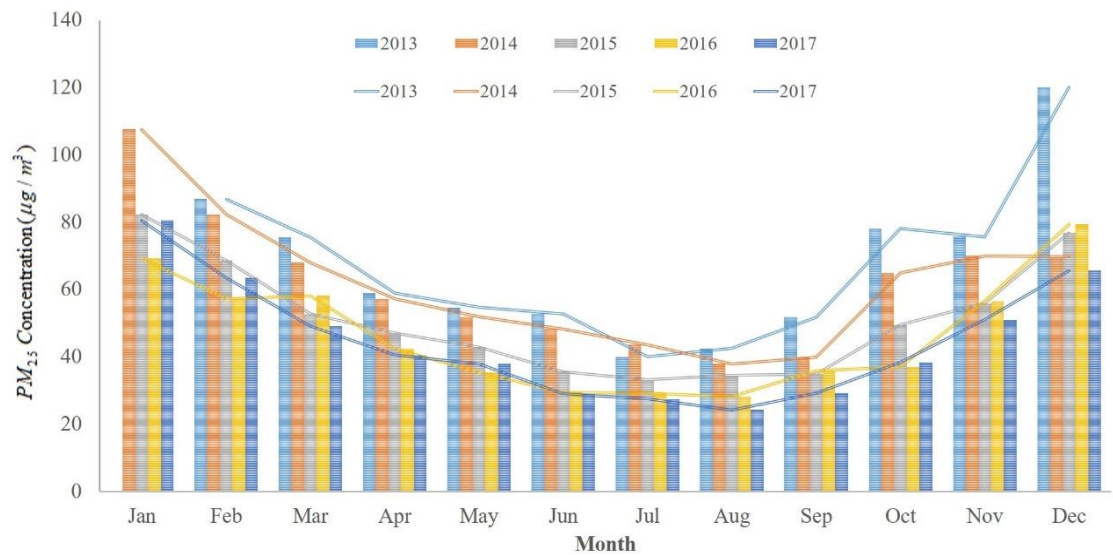

**Fig. 3.** Temporal variation of PM$_{2.5}$ concentration from 2013 to 2017.

Fig. 3 shows the monthly variation of China's average PM$_{2.5}$ concentration from 2013 to 2017. As shown in the figure, the PM$_{2.5}$ concentration in July and August is the lowest, and January and December have the highest concentrations. This means that the air pollution is much less serious in summer than in winter. This conclusion is similar to those of previous studies(Xu et al., 2017; Zhang et al., 2016a). Another phenomenon worth noting in Fig. 3 is that the position of the light blue line is at the top overall, while the dark blue line is at the bottom, which suggests that the PM$_{2.5}$ concentration of China has been decreasing from 2013 to 2017. This may indicate that the environmental protection policies and haze control measures in China have taken effect. Although the average air quality improved from 2013 to 2017, the number of cities that reached the Chinese Ambient Air Quality Standards (CAAQS) Grade I (15 $\mu g/m^3$) and Grade II standard (35 $\mu g/m^3$) is still small, and the numbers are 2 and 73 for 2015, 3 and 104 for 2016, 7 and 117 for 2017 out of the 368 cities, more than half of the cities haven't reached the Grade II standard.

We then analyzed the spatial distribution pattern of PM$_{2.5}$ concentration in China and the result of 2017 is shown below. The spatial distribution pattern has not changed much from 2013 to 2017 and has a dual-core distribution, with BTH area and Xinjiang being the two cores of high PM$_{2.5}$ concentration. Fig. 4(a) shows all the sites and the corresponding PM$_{2.5}$ concentration values we used to implement EBK in 2017, and Fig. 4(b) is the final interpolation result. The BTH region and Urumqi suffer from serious PM$_{2.5}$ pollution, but the air quality in Tibet, Yunnan province, Hainan Island, and the PRD area is better. The interpolation result is consistent with the results of previous studies(He and Huang, 2018; Li et al., 2017b; Xue et al., 2017).

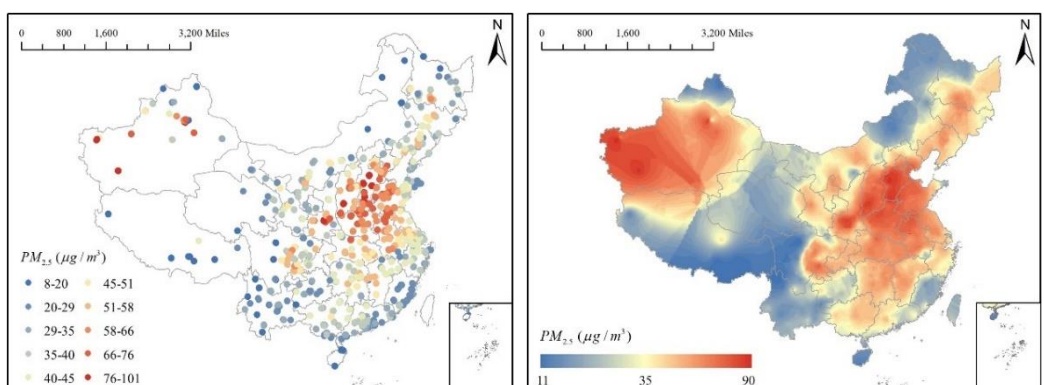

**Fig. 4.** (a) Annual PM$_{2.5}$ concentration at the environmental monitoring stations in 2017. (b) The distribution of annual average PM$_{2.5}$ concentration in 2017 in China interpolated by EBK.

*3.2. Spatial variations of the relationship*

*3.2.1 Relationship analysis at the city scale*

1) *Statistical information and general description*

We first calculated the Pearson correlation coefficients between PM$_{2.5}$ concentration and MODIS AOD in the 368 cities. All the 5-years' data are used to explore the spatial variations. The correlation coefficients r ranges from 0.013 to 0.876 with a mean value of 0.378 and a standard deviation of 0.145. We can detect obvious spatial difference from the results: the correlation between PM$_{2.5}$ concentration and AOD is high in Sichuan, Chongqing and BTH area, but the correlation is relatively lower in coastal areas such as PRD and YRD areas, with PRD region holds the lowest correlation in the table. Then we calculate the ratio of PM$_{2.5}$ concentration and AOD using the 5-years average value for each city. The ratio for 368 cities ranges from 33.031 to 471.457 with a mean value of 114.907 and a standard deviation of 62.916. The result also shows obvious spatial variations with high ratio in BTH area and relatively low ratio in PRD and YRD areas.

2) *Intuitive display and detailed analysis*

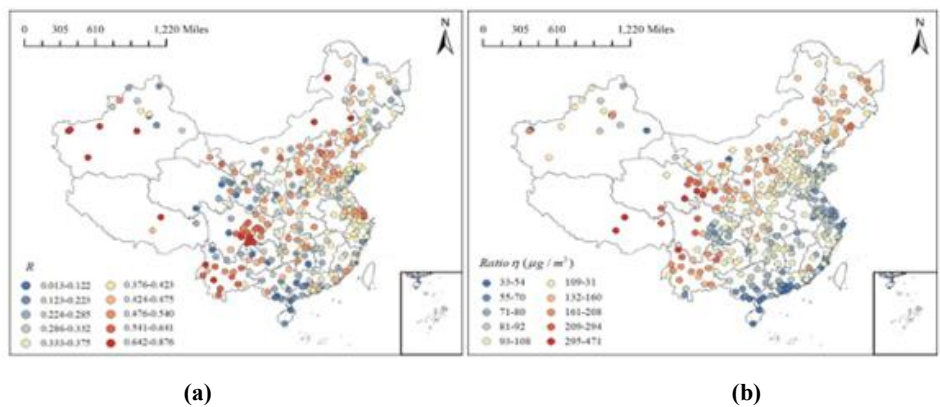

(a)  (b)

**Fig. 5.** The correlation coefficient (R) values and ratio ($\eta$) distribution in the 368 major cities

To describe the spatial variations in a more intuitive way, we display the correlation coefficients and ratios of each city on the China map in Fig. 5 and represent the correlation coefficients (r) and ratio values ($\eta$) with graduated color. The red color represents a high value and the blue color represents a low value. Fig. 5(a) shows the correlation coefficients between $PM_{2.5}$ concentration and AOD in the 368 cities, the red color mostly concentrated in eastern Sichuan, Chongqing, Yunnan Province, BTH region and some cities in southern Xinjiang, which stand for a strong correlation between $PM_{2.5}$ concentration and AOD in these areas. In contrast, the correlation in PRD areas, Hainan Province and some cities in Qinghai and Gansu Province is quite lower. It can be inferred that topography and climate have a great impact on the $PM_{2.5}$-AOD relationship. For instance, in Sichuan Province, the correlation is high in Mideastern Sichuan Basin but low in the western Sichuan plateau, the dividing line of high and low correlation is very close to the topography boundary of Sichuan Province which can be detected in the satellite image base map. Besides, In Xinjiang Province, the correlation in South is high and the correlation in north relatively weaker. The dividing line between high and low correlation is very close to the temperature and climate zone boundary.

Fig. 5(b) shows the result of ratio distribution, different from the results of r distribution, there is a conspicuous south-north difference in the figure with higher $PM_{2.5}$/AOD ratio in the north and lower ratio in the south except that Yunnan Province and western Sichuan holds a relatively high ratio but locates in the south. The north-south difference of the ratio may arise from the humidity difference in the southern and northern China. In the southern part of China, the relative humidity is usually higher than north. When the environment humidity is high, particles tends to contain more water. AOD represents the extinction ability of aerosol, so the water contained in the particles will contribute a lot to AOD. However, the concentration of $PM_{2.5}$ is dry mass concentration, the water contained in the particles will be evaporated and contribute little to $PM_{2.5}$ mass concentration in the measurements. That's to say, high humidity makes particles containing more water, thus makes AOD higher but the impact on the concentration of $PM_{2.5}$ is relatively weaker, resulting a lower $PM_{2.5}$/AOD ratio in the south.

3) *Combined analysis of correlation and ratio*

The distribution of correlation coefficients and ratio are different. The two most significant difference happen in Qinghai-Gansu-Ningxia region and Sichuan Province. In Qinghai-Gansu-Ningxia region, the correlation between $PM_{2.5}$ concentration and AOD is weak but the ratio is high, we infer that this may be the result of aerosol properties difference. We compare the Fine Mode Fraction (FMF) and Angstrom Exponent (AE) in 5 typical AERONET sites and the monthly variations of FMF and AE is shown in Fig.6. The FMF and AE value are both much lower at the site in Qinghai Province (Mt_WLG), which stands for a large proportion of coarse particles. As a previous study showed, in Qinghai-Gansu-Ningxia region, soil dust arising from frequent sand storm is a main source of $PM_{2.5}$, but not a main contributor for aerosol extinction ability and AOD, making the correlation between $PM_{2.5}$ and AOD low. High concentration of coarse particles makes $PM_{2.5}$ concentration larger but increase AOD little, thus making the ratio of $PM_{2.5}$/AOD higher. The larger proportion of coarse particles may also be able to explain the relatively lower correlation in coastal areas where coarse particles (sea salt) account for a large part in $PM_{2.5}$. But unlike Qinghai-Gansu-Ningxia region, the ratio is still low in coastal areas, that means the impact of humidity is larger than the impact of aerosol size on $PM_{2.5}$/AOD ratio in coastal areas.

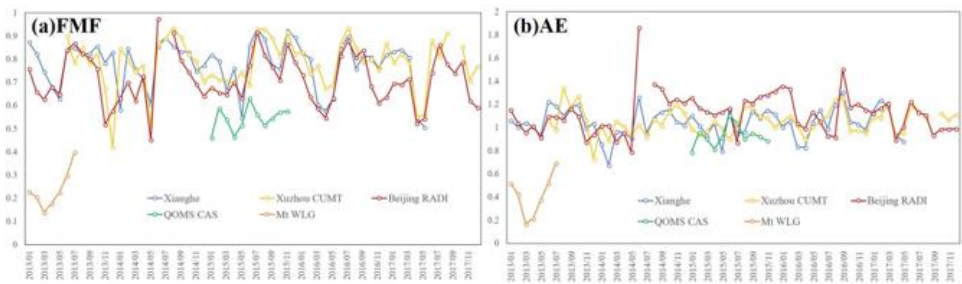

**Fig. 6.** Fine mode fraction (FMF) and Angstrom exponent (AE) monthly variations in five AERONET sites.

The second significant difference happens in Sichuan Province. In Sichuan Basin, the $PM_{2.5}$ pollution mostly comes from local industrial and vehicle emissions, the contribution of regional transportation is small, $PM_{2.5}$ and AOD tend to have the similar sources and contributors in the relatively closed region, and therefore have a higher correlation. The small ratio may result from the humid climate in Sichuan Basin.

Besides these differences, the correlation and ratio show similar distribution in most of other cities. Most typically, the correlation and ratio between $PM_{2.5}$ and AOD are both low in cities in South China. High humidity and low FMF bring about the low ratio and low correlation respectively. As for citied in the North China Plain, the correlation and ratio are both at a high level, the dry climate and serious city-industrial pollution, high FMF lead to the results.

*3.2.2 Relationship analysis at the region scale——study of 9 typical metropolitan regions*

Metropolitan regions hold an important strategic position in China's economic development, pollution research aimed at metropolitan regions is of great significance and has received extensive attention. To obtain an exact evaluation of the spatial variations of the relationship between $PM_{2.5}$ concentration and AOD at metropolitan region scale, we conducted a correlation analysis at 9 typical national metropolitan regions in China. The scatter plots for $PM_{2.5}$ concentration and AOD in 9 metropolitan regions are shown in Fig.7 Chengyu and Jingjinji metropolitan region have the highest correlation with a correlation coefficient r=0.584(CY) and

r=0.551(BTH), in consistent with the correlation analysis results at city scale. Shandong Peninsula has a correlation coefficient of 0.423 following the CY and BTH regions. YRM region has the lowest correlation among the 9 metropolitan regions. As for other metropolitan regions, the correlation coefficients are all within 0.368 to 0.392, very close to the average correlation coefficient of all cities calculated in the last section. We can notice that the results at region scale have a small difference from the results at city scale. In city scale analysis, cities in PRD regions have the lowest correlation, but the correlation turns higher in region scale analysis. That's because correlation at region scale is not a simple average of all the cities in the region, the connections and relevance between these cities can affect the correlation at region scale. The fact that the correlation between $PM_{2.5}$ and AOD of PRD region turns larger than the correlation of cities in it indicates cities in PRD region have very similar pollution condition, the correlation between cities is high, therefore, making the regional correlation higher than correlation at city scale.

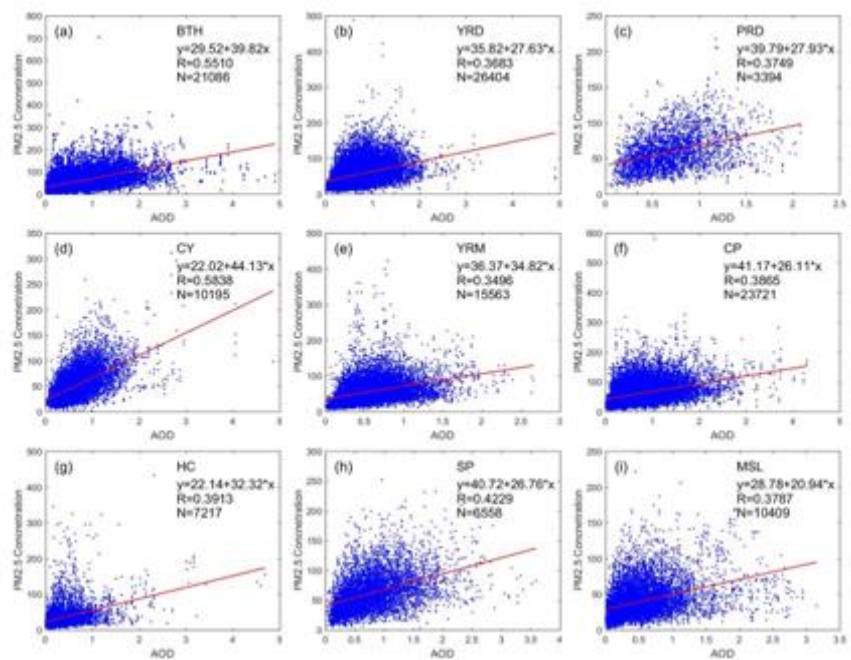

**Fig. 7.** Scatter plots for $PM_{2.5}$ concentration and AOD in the 9 metropolitan regions (The unit of $PM_{2.5}$ concentration is $\mu g/m^3$ ).

The results of ratio calculation are calculated as well, with BTH region always hold the highest ratio expect for 2015 in which Hachang metropolitan region has the highest ratio among the 9 regions. BTH area has an average ratio of 201.51 from 2013 to 2017 with HC and MSL area following behind and the ratio is 178.94 and 146.30 respectively. All the 3 metropolitan regions that have a high ratio are in the north China which is consistent with the results at city scale. The ratio in YRD, PRD, CY and YRM areas are all below 100 which is a quite low level. As for CP and SP areas, the ratio is at a middle level with $\eta=138.82$ and $\eta=110.91$.

*3.3. Temporal variations of the relationship*

*3.3.1 Monthly variations*

The relationships between $PM_{2.5}$ concentration and AOD not only vary with the geographic locations, but also with the time in a year or between years. Thus, we conducted a relationship analysis at different time scales to explore the temporal variations.

First, we calculated the correlation coefficients for each month in 9 metropolitan regions from February 2013 to December 2017 and the results are shown in Fig.8. We conduct a significance test for each correlation coefficient and the r will be display only when the P values are smaller than 0.05. Because AOD are usually missing in winter due to cloud/snow cover or high surface reflectance(Xiao et al., 2017), the correlation coefficients in cold season are mostly lacked (small number of samples makes the significance test hard to pass). The periodic variations of r are not obvious. But we can still find that different regions tend to share the similar monthly varying pattern every year. For example, in 2016 the monthly varying curves are bimodal for most regions which means there are two high correlation points in one year. One is in May and the other in September. Besides, the correlation in February (the starting point of the curve in 2016) is high too. The circumstance is similar in 2014. As for 2013, the monthly varying curve is unimodal for most regions expect for BTH, YRD and PRD. The high correlation appears in August or September. And in 2017, high correlation tends to appear in September and then comes to the trough of wave in November. The variations of the first half year are weak and the curve is nearly smooth. As we can see, the monthly varying pattern is different for each year, and though in the same year, the varying pattern may be different in some regions. On the whole, the monthly variations of r are quite complex and we cannot just come to an easy conclusion like the correlation is high in spring and low in winter. But in the terms of statistic, there is a higher probability that high correlation will appear in May and September.

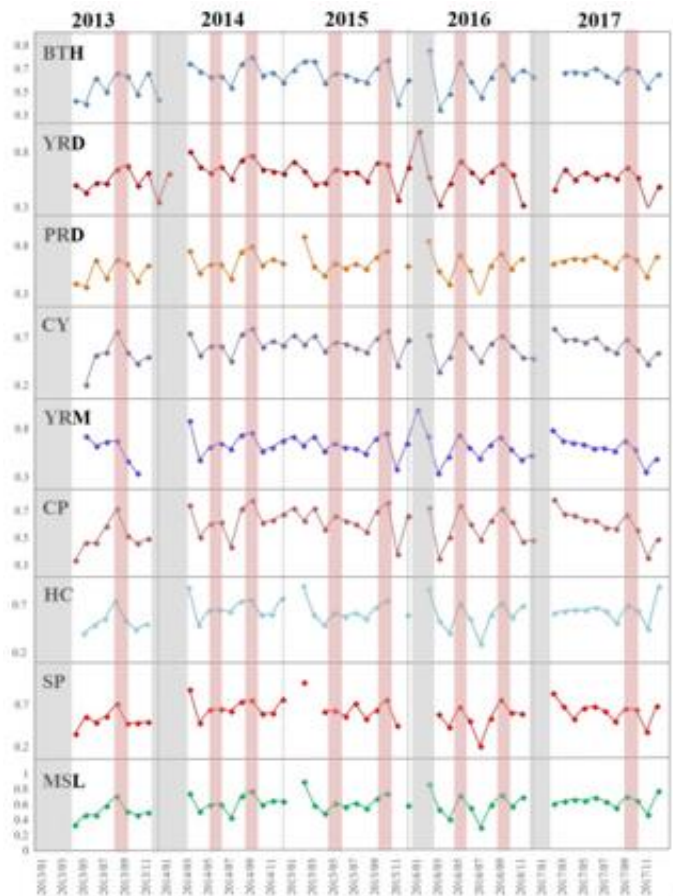

**Fig. 8.** The monthly variations of correlation coefficients r in 9 metropolitan regions. (Grey bars mark the data missing moths and the red bars highlight the high correlation months)

Then, we calculated the ratios in 9 metropolitan regions and the results are shown in Fig.9. Ratio calculation has a lower requirement for samples number than correlation analysis, thus the AOD winter missing problem has a minor impact on it. There are obvious periodic variations that the ratio is usually lower in June, July and August (warm season) and becomes higher in November, December, January and February (cold season). This difference may result from the pollution and PBLH seasonal variations. In the cold season, the $PM_{2.5}$ pollution is usually more serious with a higher $PM_{2.5}$ concentration in the atmospheric aerosols. Besides, the low PBLH in winter(Zhang et al., 2016b) makes the fine particulates mostly concentrate in the lower atmosphere, thus the surface $PM_{2.5}$ concentration, which is measured by the ground sites and studied in our paper, can account for a higher ratio in the $PM_{2.5}$ concentration for whole atmospheric column. Hence, the $PM_{2.5}$/AOD ratio can be higher in winter.

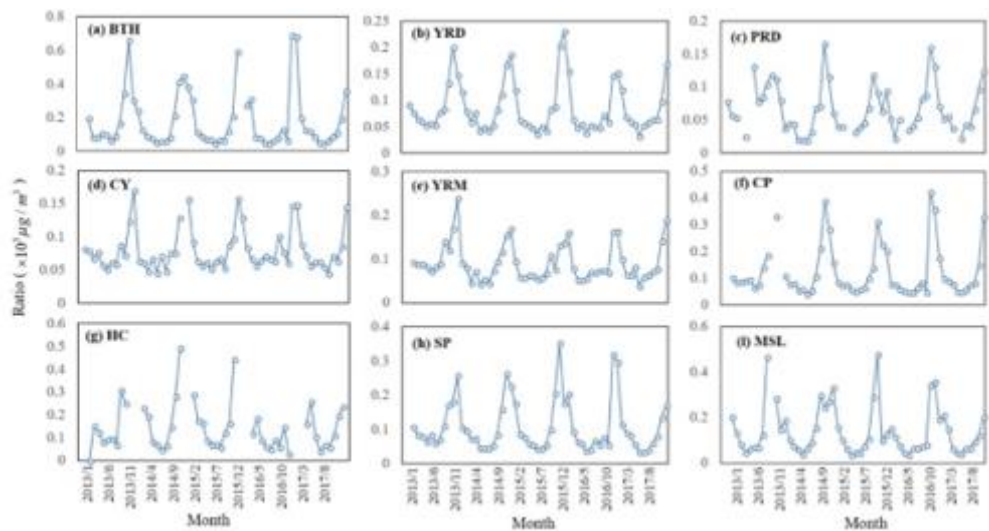

**Fig. 9.** The monthly variations of $PM_{2.5}$/AOD ratio in 9 metropolitan regions.

*3.3.2 Inter-annual variations*

In addition to the periodic analysis in the year, we also analyze the tendency variations from 2013 to 2017 to figure out how is the relationship between $PM_{2.5}$ and AOD changing in recent 5 years with $PM_{2.5}$ pollution reducing in China. Because correlation coefficient is seriously missing in cold season, and the periodic variations is not obvious, we select the highest correlation in warm season (May to December each year) in each year for 9 metropolitan regions to study the correlation interannual variations and the results are shown in Fig.13(a)and Fig.13(b) (the results for 9 metropolitan regions are shown in two pictures for better display of data). In most regions, the correlation gets stronger from 2013 to 2014 but starts to decrease since 2014.

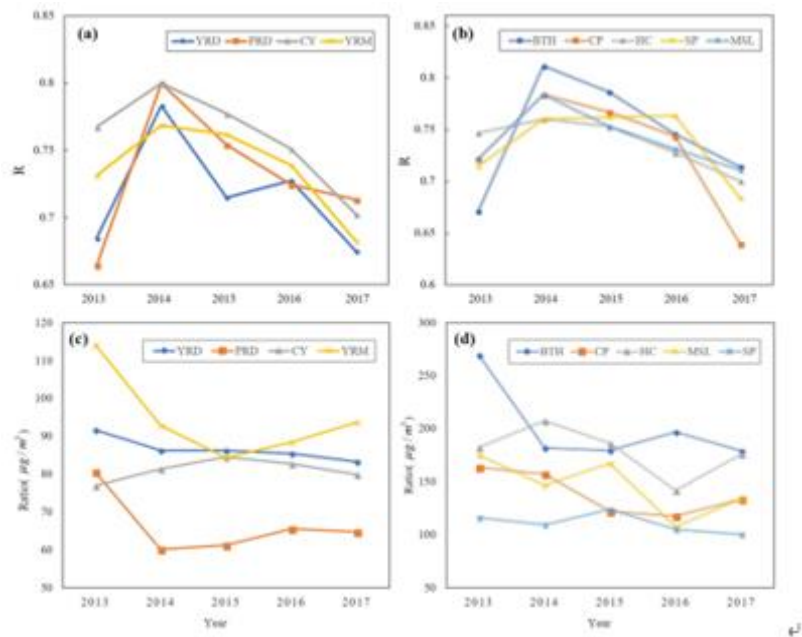

**Fig. 13.** The interannual variations of PM$_{2.5}$-AOD correlation (a, b) and their ratio (c, d) in 9 metropolitan regions. (R is the highest correlation coefficient in warm season of each year, Ratio stands for the annual average PM$_{2.5}$/AOD ratio for each year, the dotted lines in light color in figure(c) and (d) stand for the fitted tendency lines)

PM$_{2.5}$/AOD ratio for every month is complete in a year so we calculate the annual ratio for each year to analyze the interannual variations, the results are shown in the Fig.13(c) and (d). Fig.13(c) shows the results of four metropolitan regions which have a relatively lower ratio and Fig.13(d) shows the results of other metropolitan regions with a relatively higher ratio. Overall, there is a decreasing tendency from 2013 to 2017 though with some fluctuations, which indicates PM$_{2.5}$ is accounting for a less part in AOD. The decreasing tendency is more obvious in Fig.13(d) which shows the results of metropolitan regions with relative higher ratio among all the 9 regions. To quantitatively evaluate the decreasing speed, we conduct a simple linear fitting for the ratio from 2013 to 2017 and represent the descending speed with the scope of the fitting line. We find that regions with higher ratio tends to have a faster decreasing speed, for example, the linear scope for BTH areas is -16.60 and the absolute value is the largest among the 9 metropolitan regions. The linear scope for the 3 metropolitan regions in the north which have higher ratios than other regions are all less than -7, but the scope for other regions are all larger than -5, standing for a slower descending speed. Among all the 9 metropolitan regions, Chengyu Metropolitan Region is the only region that the ratio is increasing slowly on the whole and the scope of fitting line is 0.727.

*3.3.3 Interannual variations of GWR accuracy*

MODIS AOD has been widely used for $PM_{2.5}$ concentration retrieval (Guo et al., 2017b; Jung et al., 2017; Ma et al., 2014; Xie et al., 2015), but as our study shows, $PM_{2.5}$ is accounting for a less and less part in AOD and their correlation has been decreasing since 2014, so AOD's predicting ability for $PM_{2.5}$ concentration needs to be further validated. We retrieve the $PM_{2.5}$ concentration with AOD and find the retrieval $R^2$ are decreasing from 2013 to 2017, indicating the performance of GWR model are getting worse. so it's reasonable to assume that the predicting ability of AOD for $PM_{2.5}$ are descending.

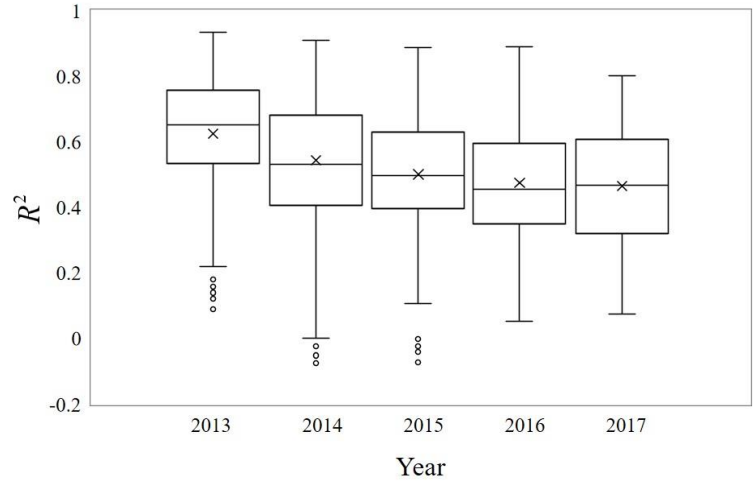

**Fig. 11.** The interannual variations of retrieval accuracy

*3.4. Discussion*

*3.4.1 Impacting factors for $PM_{2.5}$-AOD relationship*

The relationship between $PM_{2.5}$ and AOD can be influenced by many factors, through the analysis of spatial and temporal variations of $PM_{2.5}$-ADO relationship, we can simply summarize the impact of some important influencing factors.

***Planetary Boundary Layer Height (PBLH) Impact***

High PBLH makes particulate matters able to suspend in a higher vertical space, but the $PM_{2.5}$ concentration measuring instruments are usually located on the ground surface, so only the surface $PM_{2.5}$ are measured. When the PBLH gets higher, the $PM_{2.5}$ concentration we acquire from in-situ equipment will account for a less proportion in the total $PM_{2.5}$ in the atmosphere column, thus making the ratio of $PM_{2.5}$ and AOD smaller. The impact of PBLH on $PM_{2.5}$-AOD correlation is not as obvious as the impact on ratio, because PBLH mainly changes the proportion of surface $PM_{2.5}$ in total $PM_{2.5}$ but hardly changes the temporal and spatial variations pattern of $PM_{2.5}$ and AOD.

***Relative Humidity (RH) Impact***

High humidity represents a high content of water vapor in environmental atmosphere. High concentration of water vapor can contribute a lot to aerosol extinction ability and make AOD larger. But the concentration of $PM_{2.5}$ is mainly dry mass concentration, which means water vapor have a relatively small impact on $PM_{2.5}$ mass concentration. Hence, high humidity tends to make the ratio between $PM_{2.5}$ concentration and AOD smaller. Besides, due to the different impact of humidity on AOD and on $PM_{2.5}$, the existence of high concentration of water vapor may also weaken their correlation.

*Aerosol Type Impact*

Different aerosol types usually stand for different aerosol species, size and properties. City-Industry aerosol, in which sulfate and nitrate is the main species, tends to cause the high correlation between $PM_{2.5}$ and AOD. Because the extinction cross sections of sulfate and nitrate are large and they can contribution equally to AOD and $PM_{2.5}$ concentration. On the contrary, soil dust and sea salt aerosols, which usually have a larger particle size, have small extinction sections and the contribution to AOD is little compared with the contribution to $PM_{2.5}$. So, in areas with soil dust and sea salt being one of the main $PM_{2.5}$ sources, the correlation of $PM_{2.5}$ concentration and AOD tends to be weaker.

*Topography Impact*

Topography is a very important influencing factors for $PM_{2.5}$ and $PM_{2.5}$-AOD relationship, but the study of it is very lacking due to the quantitative description of topography is often difficult. In our study, we also only detect the impact of topography with a very simple and imprecise method, further research is always welcome. We find that basin terrain can usually result a high correlation between $PM_{2.5}$ and AOD in local region. For example, Sichuan Basin and Tarim Basin all have a high r value in our study. Besides, in the north part of Inner Mongolia, several cities surround by high mountains like the Great Xing 'an Mountain, Yin Mountain and Yan Mountain, high a high $PM_{2.5}$-AOD correlation as well. So, we infer that basin areas and regions surrounded by great mountains tends to form a regional pollution and $PM_{2.5}$ and AOD can share a larger account of impacting factors, thus have a higher correlation.

Finally, we should recognize that these factors are not taking effect separately and all alone but influence a city or region with a combined action. One factor may be able to explain the $PM_{2.5}$-AOD variations in one region but fails in another region. That's normal because the impacting factors have different impacting weights in different regions due to regional difference. So, when doing the analysis, we should determine the weights of different impacting factors based on local features and conduct a combined analysis.

*3.4.2 Implications for $PM_{2.5}$ retrieval through satellite AOD*

As mentioned before, AOD has been widely used to retrieve $PM_{2.5}$ concentration in all kinds of regions and periods but the performance varies a lot. In fact, AOD may not be a good indicator in some regions or with the passing by of time. In our study, we find that in coastal areas, especially lightly polluted coastal areas such as PRD region, the correlation coefficients and ratio between $PM_{2.5}$ and AOD are both small, indicating a weak linkage between $PM_{2.5}$ concentration and AOD. So, AOD may not be a very good indicator for regions like that. Besides, the interannual variations results show that the correlation between PM2.5 and AOD is getting weaker from 2014 to 2017 and $PM_{2.5}$/AOD ratio is decreasing from 2013 to 2017, the relationship between $PM_{2.5}$ and AOD is changing with time and the connections between them are attenuating in recent years. This remind us to keep a carefully attitude to AOD for $PM_{2.5}$ retrieval. In the long term, whether AOD can be a good indicator for surface $PM_{2.5}$ concentration worth deeper consideration. To sum up, with the changing of pollution condition, the role of AOD should be repositioned. In our study, we just detect the start of change, in the further future, more studies should be made to capture the essence of the change.

## 4. Conclusions

$PM_{2.5}$-AOD relationship is the cornerstone for $PM_{2.5}$ satellite retrievals. But the relationship had hardly been comprehensively and thoroughly studied for the lack of consideration for spatial and temporal heterogeneity. In this study, we analyses the $PM_{2.5}$-AOD relationship of 368 cities and 9 metropolitan regions during a period of 5 years, mainly 3 aspects of work have been done.

Firstly, we concluded the spatial and temporal variations of the relationship. Spatially, the correlation between $PM_{2.5}$ and AOD is stronger in BTH and CY, turns weaker in PRD, YRD, YRM etc. The ratio between $PM_{2.5}$ and AOD is larger in North and smaller in South. On the whole, the relationship is stronger in dry and heavily polluted regions like BTH but weak in humid and less polluted areas such as PRD. Temporally, correlation has a larger probability to be strong in May and September, monthly variations of ratio is periodic and the ratio tends to be large in cold season and small in warm season. Correlation and ratio between $PM_{2.5}$ and AOD is getting smaller in recent years.

Secondly, we summarize the impact of some influencing factors for $PM_{2.5}$-AOD relationship through the analysis of multiple data. High PBLH makes $PM_{2.5}$/AOD ratio smaller; Relative humidity weakens the correlation and decreases the ratio; Soil dust and sea salt aerosols often makes the correlation weaker and the ratio larger; Basin topography tends to improve the correlation. These factors take effect with a varying impacting weight in different regions, the combined action results the temporal and spatial variations.

Third, we conduct a simple $PM_{2.5}$ retrieval experiment and raise some suggestions for $PM_{2.5}$ satellite retrievals. The retrieval performance represented by adjusted $R^2$ of GWR is decreasing from 2013 to 2017 while $PM_{2.5}$/AOD ratio is decreasing. So, we propose that we may need to reposition the role of AOD for $PM_{2.5}$ retrieval.

Although lots of work has been done in this research, there is still much room for improvement. Firstly, the study of $PM_{2.5}$-AOD impacting factors is not comprehensive and intensive enough, studies exploring the inner physical mechanism would be able to explain the variations more thoroughly. Secondly, there are still some phenomena that we cannot fully explain, and for which further research is needed. Such as the high correlation and ratio in Yunnan Province. Further exploration of the cause of this spatial and temporal variation would help us to better understand the air pollution problem in China. Despite these limitations, our work could still provide some useful instructions for a more accurate retrieval of $PM_{2.5}$ concentration and help with improving our understanding on $PM_{2.5}$ pollution.

**Acknowledgments**

This work was supported by the National Key R&D Program of China (no. 2016YFC0200900). We would like to thank the $PM_{2.5}$ data providers of the China National Environmental Monitoring Center (CNEMC). The MODIS AOD were obtained from the NASA Langley Research Center Atmospheric Science Data center (ASDC). The NCEP/NCAR Reanalysis data are provided by the NOAA/OAR/ESRL PSD. The MERRA-2 data A were provided by the NASA GES DISC. Finally, we also thank the AERONET team for providing the ground-based aerosol data.

**Author Contributions**

Qiangqiang Yuan conceived and designed the study; Qianqian Yang analyze the data and wrote the paper; Linwei Yue and Tongwen Li contributed data collection and experimental instructions; All the authors contribute to the editing and reviewing of the paper.

**Conflicts of Interest**

The authors declare no conflict of interest.

17(21), 13473-13489.